\documentclass[12pt]{article}
\newcommand{\journal}[4]{{#1~}{#2}\,(#3)\,#4.}

\newcommand{\jmp}{\journal {J. Math. Phys.}}

\newcommand{\cmp}{\journal {Commun. Math. Phys.}}

\newcommand{\np}{\journal {Nucl. Phys.}}
\newcommand{\pl}{\journal {Phys. Lett.}}
\newcommand{\mpl}{\journal {Mod. Phys. Lett.}}
\newcommand{\prep}{\journal {Phys. Reports}}
\newcommand{\ptp}{\journal {Progr. Theor. Phys.}}

\newcommand{\annp}{\journal {Ann. Phys. }}

\newcommand{\PR}{\journal{Phys. Rev.}}
\newcommand{\jp}{\journal{J. Phys.}}

\def\text#1{\mbox{#1}}
\def\equ#1{(\ref{#1})}

\def\be#1{\begin{equation}\label{#1}}
\def\ee{\end{equation}}
\def\equ#1{(\ref{#1})}

\def\cal{\mathcal}

\def\inner#1#2{\left(#1,#2\right)}
\def\cor#1#2{\left<#1#2\right>}
\usepackage{amsmath}

\begin{document}
\begin{titlepage}
\title{{Linking number from a topologically massive p-form theory}}
 \author{M\' arcio A. M. Gomes\thanks{Electronic address: gomes@fisica.ufc.br} and R. R. Landim\thanks{Electronic address: renan@fisica.ufc.br}\\
Universidade Federal do Cear\'{a} - Departamento de F\'{\i}sica \\
C.P. 6030, 60470-455 Fortaleza-Ce, Brazil}

\end{titlepage}
 \maketitle
  \begin{abstract}
  
We  show that the linking number of two homologically trivial disjoint
$p$ and $(D-p-1)$-dimensional submanifolds of a $D$-dimensional manifold can be derived from
the topologically massive $BC$ theory in low energy regime.

\end{abstract}
\vspace{1.0cm}


\vspace{0.3cm}


Antisymmetric tensor fields arise in string theory \cite{green}
and supergravity \cite{nieu} and play an important role in
dualization \cite{spa,spa1,wots}. They can be viewed as the
components of a p-form field $B$ given by
\be{p-form} B=\frac{1}{p!}B_{\mu _{1}...\mu _{p}}dx^{\mu
_{1}}...dx^{\mu _{p}}. \ee The theory involving a $p$-form field
$B$ and a $(D-p-1)$-form field $C$ was first introduced by
Horowitz \cite{hor1} and Blau and Thompson \cite{blau1}.
Horowitz's theory does not involve any local dynamics. He was in
fact interested in generalizing  Witten's idea \cite{wit} -- who
proved the equivalence between the three dimensional Einstein
action and the non-abelian Chern-Simons term -- to an arbitrary
dimension. Horowitz treated a class of models that are invariant
under diffeomorphism, and that naturally bring ``three dimensional
gravity included as a special case''. In \cite{hor2}, Horowitz and Srednicki 
used the same model to provide a definition of generalized linking number of
$p$-dimensional and $\left( D-p-1\right)$-dimensional surfaces in
a $D$-dimensional manifold. Later, making use of variational
method, Oda and Yahikozawa \cite{oda0} obtained the same result
and generalized it to the nonabelian case.

The introduction of dynamical terms for a $p$-form field $B$ and a
$(D-p-1)$-form field $C$ leads to topologically massive theories
for abelian \cite{oda1} and non-abelian \cite{oda2} gauge
theories. These theories are a generalization of the topological
mass generation mechanism in three dimensions proposed by  Deser,
Jackiw and Templeton with the Chern-Simons term \cite{jackiw}.
This also generalizes the abelian topological mass mechanism in
$D=4$ constructed with a $2$-form and a vector field with a $BF$
term \cite{lah}. We emphasize here that the non-abelian
construction proposed in \cite{oda2} does not describes a
topolocally massive $BF$ model in $D$ dimensions. The authors did
 not include the Yang-Mills term, since they consider a flat
connection. The non-abelian topological massive Yang-Mills theory
with no flat connection was constructed in \cite{hwang} and
\cite{landim1}, in four and $D$ dimensions, respectively.

In this paper we analyze the local effects in the correlation function $\left<B(x)C(y)\right>$ of the topologically massive abelian $BC$ model integrated over two homologically trivial disjoint submanifolds.  We show that the linking number can be derived from the topologically massive $BC$ theory, in the low energy regime generalizing in part the results in \cite{hor2} and extending to $D$ dimensions the $3$-dimensional case \cite{sorella4}.

We follow closely the notation and conventions adopted in
\cite{landim2}. We use the form representation for fields with the
usual Hodge $*$ operator, which maps a $p$-form into a $(D-p)$-form and $\ast\ast=(-1)^{p(D-p)+1}$. The
 adjoint operator acting in a $p-$form is defined as $d^{\dagger
}=(-1)^{Dp+D}*d*$ \cite{nakahara}, where $d=dx^\mu (\partial
/\partial x^\mu )$ is the exterior derivative and $D$ is the
dimension of a flat manifold $\cal{M}_D$ without boundary with metric $g_{\mu
\nu }=\mbox{diag}(-++\cdots +++)$. The inner product of two $p$-forms fields $A$ and $B$ are defined by
\be{inner}
\inner{A}{B}=\int A(x)\wedge\ast
B(x)=\int_M\frac{1}{p!}A(x)_{\mu_1..\mu_p}B(x)^{\mu_1..\mu_p}d^Dx.
\ee
The $\ast d$ operator maps a $p$-form into a $(D-p-1)$-form and has the properties
\begin{eqnarray}
&&\left(\Omega_p,\ast d\Omega_{D-p-1}\right)=(-1)^{Dp+1}(\Omega_{D-p-1},\ast d\Omega_p),\label{ast1}\\
&&\left(\Omega_p,\ast d \ast d\omega_p\right)=\left(\omega_p,\ast d \ast d\Omega_p\right)\label{ast2},
\end{eqnarray}
for any $p$ and $(D-p-1)$-form.
We use from now on the rules to forms functional calculus
developed in \cite {marcio}:
\be{AAfunc} \frac{\delta A(x)}{\delta A(y)}=\delta^D_p(x-y),
\ee 
with $\delta^D_p(x-y)$ is defined in terms of usual Dirac delta function:
\begin{eqnarray}
\delta^D_p(x-y)=\frac{1}{p!}\delta^D(x-y)g_{\mu_1\nu_1}..g_{\mu_p\nu_p}dx^{\mu_1}
\wedge..\wedge dx^{\mu_p}\otimes dy^{\nu_1}\wedge..\wedge dy^{\nu_p}.
\end{eqnarray}
The linking number between two disjoint  submanifolds of $\cal{M}_D$ can be defined as
\be{link}
L\left( U,V\right)=\int_{U}\int_{W}\ast \delta _{p}^{D}\left(
x-y\right),
\ee
where $U$ and $V$ are boundaries of submanifolds $Z$ and $W$, namely, $U=\partial Z$ and $V=\partial W$. In this expression, $x$ and $y$ are points of $U$ and $W$ respectively, and the $\ast$ operator acts on the part of $\delta _{p}^{D}\left(x-y\right)$ defined on $W$.

We start with the following classical abelian action \cite{oda2},

\be{abel-act}
S=\int_{\cal{M}_{D}}\left( \frac{1}{2}(-1)^r H_{B}\wedge \ast H_{B}+\frac{1}{2}%
(-1)^s H_{C}\wedge \ast H_{C}+mB\wedge dC\right), \ee where
$r=Dp+p+D$,  $s=Dp+p+1$, $B$ is a $p$-form field, $C$ is a
$(D-p-1)$-form field both with canonical dimension $(D-2)/2$ and
$H_{B}$, $H_{C}$ are their respective field strengths

\begin{eqnarray}
H_{B} &=&dB,  \label{HB} \\
H_{C} &=&dC, \label{HC}
\end{eqnarray}
all them real-valued and $m$ is a mass parameter. The factor
$(-1)$ in front of the kinetic terms is required in order to have
a positive kinetic energy in the Hamiltonian. As claimed in \cite{oda2}, the model just describe a topologically massive $BC$ model denoted by $TMBC$. Note that for $D=4$
and $p=1$ we recover the topologically massive $BF$ model
\cite{lah}. The action is clearly invariant under the gauge
transformations
\begin{eqnarray}
\delta B &=&d\Omega,    \label{deltaB} \\
\delta C &=&d\Theta,    \label{deltaC}
\end{eqnarray}
where $\Omega $ and $\Theta $ are $(p-1)$-form and $(D-p-2)$-form
gauge parameters. These gauge transformations are reducible,
\textit{i.e.},$\Omega ^{\prime }$ and $\Theta ^{\prime }$ given by

\begin{eqnarray}
\Omega ^{\prime } &=&\Omega +d\omega,    \label{omega} \\
\Theta ^{\prime } &=&\Theta +d\theta,     \label{theta}
\end{eqnarray}
 are also honest gauge parameters satisfying  \equ{deltaB} and \equ{deltaC}
respectively, since $d^{2}=0$. Naturally, the same holds to $\omega $, $%
\theta $, etc. So, in order to construct the action to be
quantized, one has to introduce ghosts and ghosts for ghosts and
so on.

Let us write the action in a more compact form. We introduce a doublet $\Phi(x)$, with $B(x)$ and $C(x)$ being the components fields:
\be{dou}
\Phi(x)= \left(
\begin{array}{c}
\Phi_1(x) \\
\Phi_2(x) \\
\end{array}
\right) = \left(
\begin{array}{c}
B(x) \\
C(x) \\
\end{array}
\right).
\ee

The inner product between two doublets  is defined by
\be{inner2}
\inner{\Phi}{\Psi}=\inner{\Phi_1}{\Psi_1}+\inner{\Phi_2}{\Psi_2}=\inner{\Psi}{\Phi}.
\ee
Then, making the use of Eqs. \equ{ast1} and \equ{ast2}, we have
\be{action-red}
S_{TMBC}=\frac{m}{2}\inner{\Phi}{\ast^{-1} d\cal{O}\Phi},
\ee
where 
\be{opr}
\cal{O}=
\left(
\begin{array}{cc}
(-1)^{Dp+D+1}\ast d/m & 1 \\
(-1)^{Dp+1}&(-1)^{Dp+D+1}\ast d/m\\
\end{array}
\right).
\ee
We are interested in the computation of 
\be{corr}
\left<\int_U B(x)\int_V C(y)\right>_{TMBC}.
\ee
In order to obtain this correlation function, we must deal with the gauge-fixed action.

The gauge fixed action becomes,
\begin{equation}\label{SGF}
S_{gf} =\frac{m}{2}\inner{\Phi}{\ast
^{-1}d\cal{O}\Phi} +\inner{L}{d\ast\Phi}
+\ldots,
\end{equation}
where the dublet
\begin{equation}\label{dubNL}
L(x) = \left(
\begin{array}{c}
L_{1}(x) \\
L_{2}(x) \\
\end{array}
\right),
\end{equation}
is the Nakanishi-Lautrup field introduced to implement the evaluation of path integral. Note that $L_{1}$ and $L_{2}$ are a $(D-p+1)$-form and a
$(p-2)$-form respectively. The functional is written as
\be{fun}
Z=\int \cal{D}Xe^{iS_{gf}},
\ee
where $\cal{D}X=\cal{D}B\cal{D}C\cal{D}L_1\cal{D}L_2\cdots$ is the functional measure.
From the functional identities
\be{zphi}
\frac{1}{Z}\int \cal{D}X\frac{\delta }{\delta \Phi \left(
y\right) }\left[ \Phi \left( x\right) e^{iS_{gf}}\right] =0,
\ee
and
\begin{equation}
\frac{1}{Z}\int \cal{D}X\frac{\delta }{\delta \Phi \left(
y\right) }\left[ L\left( x\right) e^{iS_{gf}}\right] =0,
\end{equation}
we have
\be{dubid2} im\left< \Phi \left( x\right) \ast
^{-1}d\cal{O}\Phi \left( y\right) \right> \pm i\left< \Phi
\left( x\right) d\ast L\left( y\right) \right> +\delta
_{p,D-p-1}^{D}\left( x-y\right) =0, \ee 
 \be{zL}\left< L\left( x\right) \ast^{-1}d \cal{O}
\Phi\left( y\right) \right>  \pm \left< L \left( x\right) d\ast
L\left( y\right) \right> = 0,
\ee
where
\begin{equation}\label{dubdelta}
\delta _{p,D-p-1}^{D}\left( x-y\right)= \frac{\delta \Phi \left(
x\right) }{\delta \Phi \left( y\right) }= \left(
\begin{array}{c}
\delta _{p}^{D}(x-y) \\
\delta _{D-p-1}^{D}(x-y) \\
\end{array}
\right),
\end{equation}
and the correlation function of two dublets is taken as being 
\be{cor-dub}
\cor{\Phi(x)}{\Psi(y)}=\left(
\begin{array}{c}
\cor{\Phi_{1}(x)}{\Psi_1(y)} \\
 \cor{\Phi_{2}(x)}{\Psi_2(y)}\\
\end{array}
\right).
\ee
To compute these correlation functions, we must invert the operator $\cal{O}$. But $\cal{O}^{-1}$ has local and non-local terms. To get rid of non-local terms, one has to be concerned with low energy regime. So, to get the local terms of $\cal{O}^{-1}$, we expand in powers of $\ast d/m$:
\be{inverseO}
\cal{O}^{-1}=
\left(
\begin{array}{cc}
(-1)^{D+1}\ast d/m & (-1)^{Dp+1} \\
1&(-1)^{D+1}\ast d/m\\
\end{array}
\right)\alpha,
\ee
where 
\be{alpha}
\alpha=\Sigma_{n=0}^\infty(-1)^{n(Dp+1)}(\ast d/m)^{2n}.
\ee
Since $L$ is Nakanishi-Lautrup field, $\cor{L(x)}{L(y)}=0$. Then, from  Eq. \equ{zL}, we have
\be{ldphi}
\cor{d\Phi(x)}{L(y)}=0,
\ee
and consequently,
\begin{eqnarray}
\left<\int_U B(x)\int_W \ast d\ast L_1(y)\right>=0.
\end{eqnarray}
Using this identity and Eq. \equ{dubid2} we arrive at
\be{intuw}
m\cor{\int_U B(x)}{\int_V C(y)}+(-1)^{Dp+D+1}\cor{\int_U B(x)}{\int_V \ast dB(y)}=iL(U,V),
\ee
where we have used the Eq. \equ{link}. To evaluate the second term of the equation above, we take $x\ne y$ in the equation \equ{dubid2} and apply $\cal{O}^{-1}$ on it:
\be{local-1}
\cor{\Phi(x)}{\ast d\Phi(y)}=\frac{\pm1}{m}\cor{\Phi(x)}{\cal{O}^{-1}d\ast L(y)}.
\ee
In low energy regime $\cal{O}^{-1}d\ast=\beta d\ast$, where 
\be{beta}
\beta=
\left(
\begin{array}{cc}
0& (-1)^{Dp+1} \\
1&0\\
\end{array}
\right).
\ee
Writing Eq. \equ{local-1} in components and integrating over $U$ and $V$, it is clear that
\be{local-2}
\cor{\int_U B(x)}{\int_V\ast d B(y)}=\frac{\pm1}{m}\cor{\int_U B(x)}{\int_V d\ast L_2(y)}=0.
\ee
So, we finally have that
\be{int-link}
iL(U,V)=m\cor{\int_U B(x)}{\int_V C(y)}.
\ee
We must enforce that  this remarkable result  was deduced restricting ourselves to low energy regime. Otherwise, non-local terms would appear and could jeopardize our analysis.
\vskip1cm
  \noindent
  {\large\bf Acknowledgments}
  \vskip0.5cm

  We wish to thank J. R. Goncalves for reading the manuscript. Conselho Nacional de
   Desenvolvimento Cient\'\i fico e tecnol\'ogico-CNPq is gratefully acknowledged for financial support.
\vskip0.5cm
 {\bf Dedicatory}

   ``My wife is a great person and I love her`` (R. R. Landim).

\end{document}